\documentclass[11pt]{article}
\usepackage[margin=1in]{geometry}
\usepackage[T1]{fontenc}
\usepackage[utf8]{inputenc}
\usepackage{lmodern}
\usepackage{microtype}
\usepackage{booktabs}
\usepackage{graphicx}
\usepackage{amsmath}
\usepackage{enumitem}
\usepackage{caption}
\usepackage[hidelinks]{hyperref}
\usepackage{xurl}

\title{Revocation-Ready CP-ABE Key Management for Blockchain-Based IoT Data Sharing}
\author{CHUN YIN CHIU\\King's College London}
\date{}

\begin{document}
\maketitle

\begin{abstract}
Blockchain-based IoT data sharing systems increasingly adopt a hybrid architecture: a permissioned ledger stores tamper-evident metadata while encrypted payloads are placed in content-addressed storage, such as IPFS. In such systems, a dominant security bottleneck is key access control: enforcing dynamic, multi-user authorization for the bulk-data decryption key. Existing designs often rely on always-online RBAC or smart-contract gates that return keys to authorized users, reintroducing a trusted online policy enforcement point and weakening auditability. We present a revocation-ready key management layer that replaces online key release with ciphertext key publication: the ledger records $(CID, CK, PolicyID, epoch)$, where $CK$ is a CP-ABE ciphertext encapsulating an AES-GCM key. Users retrieve $CK$ directly from the ledger and decrypt locally if their attributes satisfy the policy. To support forward revocation and policy evolution without re-encrypting large files, we introduce an epoch/time-bound attribute and a lightweight $CK$-rotation protocol that updates only small ciphertext keys and ledger entries. We implement a minimal end-to-end prototype using a local content-addressed store, a hash-chained ledger, and a CP-ABE backend; the prototype is designed to isolate key-management costs rather than benchmark production blockchain or IPFS throughput. Results show that (i) end-to-end store latency is dominated by CP-ABE encryption (approximately 186 ms for a $k=6$ mixed-Boolean policy), while ledger and storage operations are small in the local prototype; (ii) epoch-based revocation amortizes key update cost and is insensitive to churn within an epoch; (iii) under a simulated $4\times$ client slow-down, gateway-assisted mode reduces median client-side decryption time by more than $4\times$; and (iv) ledger growth scales with the number of shared assets rather than the number of readers. We provide a reproducible artifact that generates the figures and tables used in this study.
\end{abstract}


\noindent\textbf{Additional Keywords and Phrases:} blockchain, IoT data sharing, attribute-based encryption, revocation, outsourced decryption, operational cost.

\section{Introduction}
Internet of Things (IoT) deployments generate continuous streams of sensitive data -- industrial telemetry, smart-building logs, health measurements -- that must be shared with multiple stakeholders such as operators, auditors, and service providers under evolving authorization policies. A common architectural pattern is to decouple integrity and storage: a permissioned blockchain provides an immutable audit trail~\cite{fabric} for metadata, while encrypted payloads are stored off-chain in content-addressed storage such as IPFS~\cite{ipfs}. This hybrid design improves provenance and availability, but it surfaces a persistent problem: how to control access to the decryption key for each payload without reintroducing a trusted online bottleneck.

In many blockchain-based sharing proposals, key access control is implemented as an RBAC gate or smart contract that decides whether to release a symmetric key to a requester. While simple, online key release has three drawbacks. First, it creates a single policy enforcement point that must be trusted and kept online. Second, it complicates auditability: the ledger may record who asked, but not necessarily what was released. Third, it makes policy evolution expensive: revocation often implies either re-encrypting the data or maintaining per-user key state that grows with churn.

Attribute-Based Access Control (ABAC) provides a more expressive authorization model than roles~\cite{nistabac}, allowing policies to be written over user or device attributes, such as role=maintainer, site=plantA, and clearance=high. Ciphertext-Policy Attribute-Based Encryption (CP-ABE)~\cite{bethencourt} operationalizes ABAC cryptographically by embedding the access policy into the ciphertext: a data owner encrypts under a policy, and any user whose private key satisfies the policy can decrypt. This suggests an alternative to online key release: publish a ciphertext key ($CK$) on the ledger and let eligible users decrypt it locally.

However, CP-ABE is not free in practice: policy satisfaction is computationally heavier than symmetric crypto, and revocation or policy updates are non-trivial because ABE keys are long-lived. These operational concerns are particularly acute in IoT environments where endpoints may be resource-bounded.

We address these gaps with a revocation-ready CP-ABE key management layer for blockchain-based IoT data sharing. Our design keeps bulk data encrypted under AES-GCM and uses CP-ABE only to encapsulate the AES key. We add an epoch/time-bound attribute to make forward revocation and policy evolution practical: ciphertext keys are periodically rotated without touching large payloads. We further provide an optional gateway-assisted mode to reduce client cost via outsourced decryption.

We use the term blockchain-based to describe the target architecture and metadata semantics. Our prototype emulates the ledger and content-addressed storage locally to isolate cryptographic and key-management costs; it is not intended as a production Hyperledger Fabric or IPFS throughput benchmark.

\paragraph{Contributions.}
\begin{itemize}[leftmargin=*]
    \item \textbf{Design:} an auditable key-release workflow that stores $(CID, CK, PolicyID, epoch)$ on a ledger and encrypted payloads in content-addressed storage, replacing online RBAC gating with ciphertext publication.
    \item \textbf{Revocation readiness:} an epoch-based protocol for forward revocation and policy evolution that updates only small ciphertext keys and ledger entries, with a clear operational cost model.
    \item \textbf{IoT-aware client mode:} a gateway-assisted workflow that reduces client-side CP-ABE cost under standard outsourced-decryption assumptions, without giving the gateway direct access to the AES key in the intended model.
    \item \textbf{Evaluation and artifact:} a minimal prototype and reproducible evaluation suite (Experiments 1--7) executed on a commodity MacBook, including end-to-end latency, revocation cost, operational scaling, gateway trade-offs, and a case study.
\end{itemize}

\paragraph{Scope.}
Our goal is not to benchmark consensus throughput. Rather, we isolate and quantify key-management costs that persist across ledger implementations: CP-ABE encryption/decryption, $CK$ rotation under revocation, and per-asset metadata growth. This complements system-level ledger benchmarks rather than replacing them.

\section{Background and Related Work}
Hybrid on-chain/off-chain storage has become a common design choice for data-intensive blockchain applications. Content-addressed storage systems such as IPFS~\cite{ipfs} allow large encrypted objects to be referenced by a stable content identifier (CID) without storing the bytes on-chain. Permissioned ledgers such as Hyperledger Fabric~\cite{fabric} provide integrity, ordering, and access control for metadata in consortium settings and are frequently used in enterprise and IoT deployments.

Attribute-based encryption (ABE) enables cryptographic access control. In CP-ABE, access policies are attached to ciphertexts and user keys carry attributes~\cite{bethencourt}. Subsequent work improved expressiveness and efficiency~\cite{waters}. In our prototype, we use a CP-ABE construction aligned with FAME (Fast Attribute-based Message Encryption)~\cite{fame} via an open-source library~\cite{gofe}.

Revocation in ABE is challenging because ciphertexts are not tied to a single recipient and user keys may be long-lived. Broadly, revocation mechanisms fall into three families: indirect revocation via periodic key updates or attribute versioning, often combined with proxy re-encryption or lazy re-encryption of ciphertext components~\cite{yu}; direct revocation where ciphertexts include a revocation list or related mechanism~\cite{yuen}; and time-based approaches that bind decryption capability to a time period or epoch~\cite{yuen}. Recent surveys discuss the design space and IoT-specific constraints~\cite{survey}.

IoT endpoints are often resource-constrained and may not tolerate repeated pairing-heavy operations required by ABE. Outsourced decryption addresses this issue by allowing an untrusted helper to transform a ciphertext into a form that the client can finish decrypting with minimal work~\cite{green}.

Several works apply blockchain to access control and data sharing in IoT, emphasizing auditability and decentralized authorization~\cite{dorri,dai}. Many proposals, however, either assume an always-online key server or provide limited empirical evidence about revocation and operational costs. Our work targets this gap by combining a runnable minimal pipeline, a revocation-ready key update mechanism, and a cost-oriented evaluation tailored to commodity hardware.

\section{System and Threat Model}
\paragraph{System entities.} The system comprises: (1) a data owner, such as an IoT device or an organization controlling IoT data, that produces a data object and defines an access policy; (2) a content-addressed storage network that stores encrypted payloads ($CF$) and returns their CIDs; (3) a permissioned ledger that stores append-only metadata records; (4) an ABE authority that runs Setup and issues attribute keys, including epoch updates; (5) data consumers that retrieve ciphertext keys ($CK$) and encrypted payloads to obtain plaintext; and (6) an optional gateway or edge node that supports outsourced decryption for resource-bounded clients.

\paragraph{Data model.} Each data object is encrypted under a fresh symmetric key $K$ using AES-GCM~\cite{aesgcm} to produce $CF$. The key $K$ is encrypted under CP-ABE with a policy $P$ to produce $CK$. The ledger records a metadata tuple $\langle CID, CK, PolicyID, epoch, ownerID, timestamp\rangle$. The $PolicyID$ may refer to a canonical policy string or a hash thereof.

\paragraph{Adversary model.} We assume the content-addressed storage is honest-but-curious: it stores and serves bytes but may attempt to learn plaintext. The ledger is integrity-protected by its consensus protocol, or by an append-only hash-chained log in our prototype, and may be read by all participants. Network adversaries may observe ledger and storage traffic. Consumers may be malicious and attempt to decrypt $CK$ without satisfying policies. We assume the ABE authority is trusted to generate correct keys; key escrow is discussed as a limitation.

\paragraph{Security goals.} Our primary goal is confidentiality of the payload and symmetric key: unauthorized parties, including storage, should not learn plaintext data or $K$. We additionally require auditability: ledger metadata should provide a tamper-evident trail of which ciphertext keys were published and when. We focus on confidentiality and auditability; policy privacy, including hiding attributes and policies, is out of scope. Table~\ref{tab:threat} summarizes the trust assumptions and security objectives.

\begin{table}[t]
\centering
\caption{Threat model summary.}
\label{tab:threat}
\small
\begin{tabular}{llll}
\toprule
Component & Assumption & Adversary capability & Goal \\
\midrule
CAS/IPFS storage & honest-but-curious & reads $CF$; drops data & no plaintext; CID integrity \\
Permissioned ledger/log & append-only; integrity & reads metadata; tampers & auditability; tamper-evidence \\
Gateway (optional) & untrusted helper~\cite{green} & sees transform; deviates & cannot learn $K$; verifiable \\
Consumer & may be malicious & partial attributes & fails unless policy satisfied \\
ABE authority & trusted setup; escrow & issues keys/epochs & correct issuance; mitigations \\
\bottomrule
\end{tabular}
\end{table}

\paragraph{Revocation semantics.} Like most cryptographic revocation mechanisms, our approach provides forward revocation: after an epoch rollover and $CK$ update, a revoked principal can no longer decrypt newly published ciphertext keys. This cannot retroactively prevent access to plaintext that was already decrypted or to previously obtained keys; it bounds future access.

\begin{figure}[t]
\centering
\includegraphics[width=0.88\linewidth]{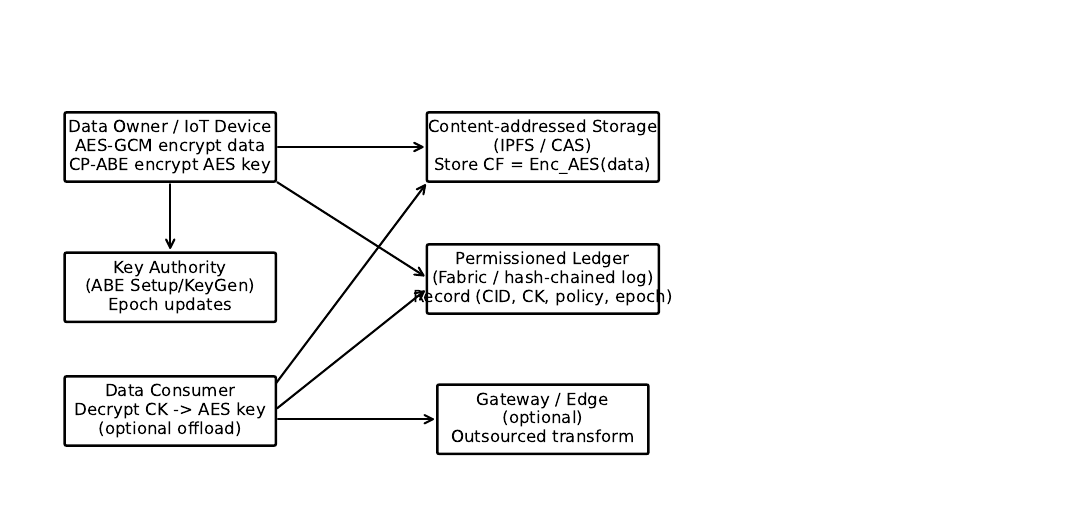}
\caption{System components and data flow in the proposed ledger plus content-addressed storage architecture. The gateway is optional.}
\label{fig:system}
\end{figure}

\section{Design}
This section describes the key-release workflow, policy representation, epoch-based revocation, and a simple cost model. The core principle is to keep bulk data encrypted symmetrically, for efficiency, while using CP-ABE only to encapsulate a short symmetric key.

\subsection{Key release via ciphertext publication}
Given a data object $D$ and access policy $P$, the owner generates a fresh AES key $K$ and computes the encrypted payload $CF = \mathrm{AES\mbox{-}GCM.Enc}(K, D)$~\cite{aesgcm}. The owner then computes a ciphertext key $CK = \mathrm{CP\mbox{-}ABE.Enc}(P, K)$~\cite{bethencourt}. The encrypted payload $CF$ is placed in content-addressed storage, returning a CID. Finally, the owner appends a metadata record to the ledger containing CID, $CK$, policy metadata, and the current epoch identifier.

To access the object, a consumer reads the ledger record, fetches $CF$ by CID, and attempts to decrypt $CK$ using its attribute key. If the consumer's attributes satisfy $P$, it obtains $K$ and decrypts $CF$. If not, decryption fails and the consumer learns nothing about $K$. Since old $CK$ versions remain on the ledger for auditability, clients must select the latest $CK$ for a given CID, for example by choosing the maximum epoch.

\subsection{Policy language}
We assume Boolean access policies over attributes, with AND and OR, consistent with CP-ABE~\cite{bethencourt,waters}. Policies encode ABAC constraints such as $(role=maintainer \wedge site=plantA \wedge clearance \geq 2)$~\cite{nistabac}. Our prototype supports conjunctions and mixed AND-of-OR forms that approximate realistic ABAC expressions while remaining compatible with standard CP-ABE toolchains.

\subsection{Revocation and policy evolution via epochs}
Revocation is difficult in ABE because ciphertexts are broadcast to attribute sets rather than specific users, and user keys may be long-lived. We adopt a time-bound epoch attribute approach: every access policy includes a special attribute $epoch=t$ for the current epoch $t$~\cite{yu,yuen,survey}.

A user's private key includes $epoch=t$ only if the authority deems the user non-revoked for that epoch. When revocation occurs, the authority advances the epoch to $t+1$ and issues updated epoch attributes only to non-revoked users.

To enforce forward revocation without re-encrypting large payloads, we rotate only ciphertext keys $CK$. For each protected object, the owner or an authorized rekey service re-encrypts $K$ under an updated policy $P'$ that includes $epoch=t+1$ and appends a new ledger record with $CK'$. The payload $CF$ and CID remain unchanged.

\subsection{Gateway-assisted outsourced decryption}
IoT endpoints may not tolerate repeated pairing-heavy operations required by ABE decryption. Outsourced decryption allows an untrusted helper to transform a ciphertext into a form that the client can finish decrypting with minimal work~\cite{green}. We support an optional gateway or edge node that performs most CP-ABE decryption work and returns a transformed ciphertext to the client. Standard constructions prevent the gateway from learning $K$ by giving it a transformation key and keeping a small client secret~\cite{green}.

\subsection{Operational cost model}
Let $M$ be the number of protected assets, $N$ the number of consumers, and $|P|$ the policy size. Let $T$ denote an observation window and let $L$ be the epoch length, so the number of epochs is $E=\lceil T/L \rceil$. Let $R$ be the number of revocation events in $T$.

Publishing a new asset requires one CP-ABE encryption and one ledger write: $O(M \cdot Cost_{ABEEnc}(|P|) + M \cdot Cost_{LedgerWrite})$. Each read requires a ledger read, a CAS fetch, a CP-ABE decryption attempt, and AES-GCM decryption: $O(Cost_{LedgerRead} + Cost_{Fetch} + Cost_{ABEDec}(|P|) + Cost_{AES})$. Naive per-revocation rekeying performs $M \cdot R$ $CK$ updates, while epoch-based rekeying performs $M \cdot E$ updates. The update work ratio is therefore $(naive/epoch)=R/E$, predicting the qualitative behavior in Experiment~2: under high churn, epoch-based rekeying amortizes work and becomes substantially cheaper.

\section{Implementation}
We implement a runnable prototype that captures the end-to-end pipeline: (i) a content-addressed store for encrypted payloads, (ii) an append-only ledger for metadata, and (iii) a CP-ABE backend for $CK$ encryption and decryption. The prototype is intentionally minimal to make operational costs visible and to support reproducibility.

\paragraph{ABE backend.} We use a Go-based CP-ABE daemon built on the gofe library~\cite{gofe}, which includes an implementation aligned with the FAME construction~\cite{fame}. The daemon exposes encrypt/decrypt commands over stdin/stdout to simplify benchmarking from Python. AES-GCM uses 96-bit nonces and 128-bit tags~\cite{aesgcm}.

\paragraph{Content-addressed store.} We implement a local CAS that stores blobs by their SHA-256 digest. This models the key performance property of systems like IPFS~\cite{ipfs}: payload retrieval is keyed by content rather than location.

\paragraph{Ledger.} We implement a hash-chained append-only log, where each record includes the hash of the previous record, providing tamper-evidence similar to blockchain ledgers. In practice, this component maps to a permissioned blockchain such as Hyperledger Fabric~\cite{fabric}; our prototype does not benchmark Fabric consensus or endorsement latency.

\paragraph{Artifact.} A single command produces all results, including CSV/JSON tables and figures. The harness records means and tail latencies, including p50 and p99 where applicable. A manifest file enumerates all generated artifacts and their paths.

\section{Prototype Evaluation}
\subsection{Experimental setup}
\paragraph{Hardware/software.} All experiments were executed on a commodity MacBook Pro-class laptop running macOS, using Python 3.13 for the orchestration harness and a Go CP-ABE daemon for cryptographic operations. The artifact installs required Python dependencies in a local virtual environment and re-executes within that environment to support reproducibility.

\paragraph{Workloads and metrics.} We report end-to-end store/retrieve latency breakdowns (Experiment~1), revocation cost under churn (Experiment~2), operational scaling with users and assets (Experiment~3), gateway offload trade-offs (Experiment~4), baseline comparisons (Experiment~5), sensitivity to ledger batching (Experiment~6), and a case study (Experiment~7). Unless otherwise noted, crypto timings reflect measured wall-clock time for CP-ABE and AES-GCM operations, and ledger/CAS timings reflect the local prototype implementation costs.

\subsection{Experiment 1: End-to-end microbenchmarks}
Experiment~1 measures the end-to-end cost of storing and retrieving an encrypted object using the full pipeline: AES-GCM for payload encryption, CP-ABE for key encapsulation, content-addressed storage operations, and ledger metadata operations. We vary file size (1 KB, 10 KB, 1 MB) and policy complexity ($k=3$ versus $k=6$, and pure AND versus mixed AND-of-OR).

\paragraph{Key insight 1: policy cost dominates store.} CP-ABE encryption cost grows with policy complexity and dominates end-to-end store latency, while symmetric encryption and storage/ledger operations remain small. For a 1 MB object encrypted under a mixed Boolean policy with $k=6$, CP-ABE encryption takes 185.9 ms out of 191.2 ms end-to-end (97.2\%). In contrast, AES-GCM encryption is approximately 1.2 ms and ledger/store costs are small in the local prototype.

\paragraph{Key insight 2: retrieve is dominated by CP-ABE decryption, but is smaller.} Retrieval is faster because it avoids CP-ABE encryption. For the same 1 MB configuration, CP-ABE decryption takes 11.1 ms out of 14.6 ms end-to-end (76.1\%), and all non-ABE steps sum to approximately 3.5 ms.

\begin{table}[t]
\centering
\caption{Experiment 1 microbenchmarks for a 1 MB object. CP-ABE encryption dominates store latency and grows with policy complexity, while retrieval remains dominated by CP-ABE decryption.}
\label{tab:micro}
\begin{tabular}{llrrrr}
\toprule
Policy form & $k$ & ABE enc (ms) & ABE dec (ms) & Store E2E (ms) & Retrieve E2E (ms) \\
\midrule
AND & 3 & 42.13 & 11.13 & 47.28 & 15.37 \\
AND & 6 & 102.60 & 11.17 & 107.36 & 16.33 \\
AND-of-OR & 3 & 67.74 & 11.22 & 72.73 & 15.61 \\
AND-of-OR & 6 & 185.90 & 11.10 & 191.20 & 14.59 \\
\bottomrule
\end{tabular}
\end{table}

\begin{figure}[t]
\centering
\includegraphics[width=0.72\linewidth]{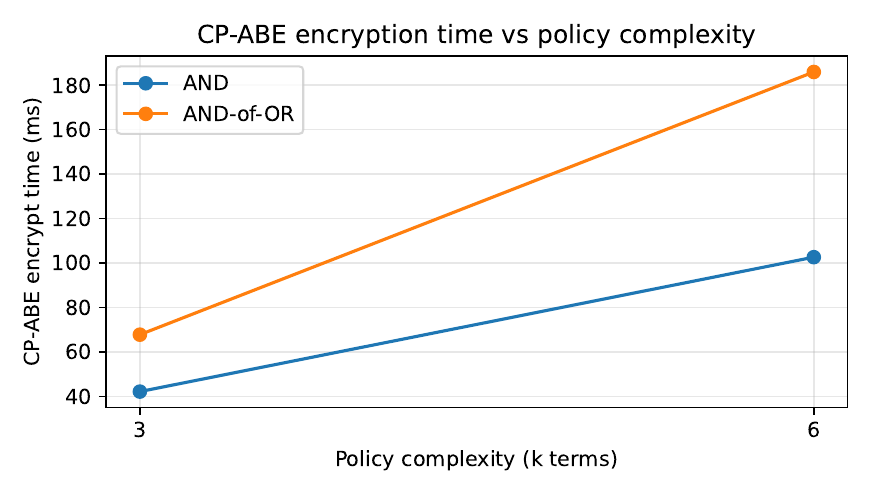}
\caption{CP-ABE encryption time versus policy size $k$ for two policy forms using a 1 MB object.}
\label{fig:policy}
\end{figure}

\begin{figure}[t]
\centering
\includegraphics[width=0.90\linewidth]{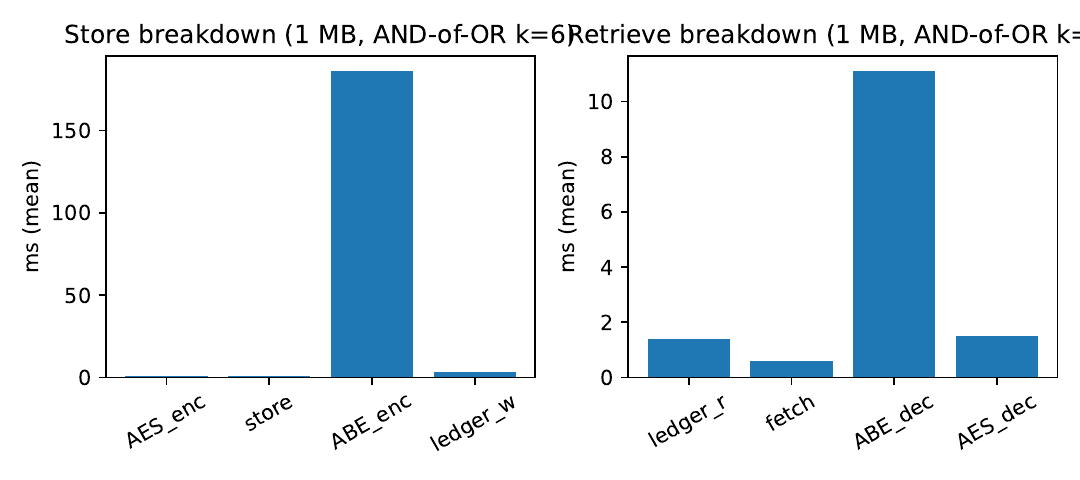}
\caption{End-to-end store and retrieve latency breakdown for a 1 MB object under a mixed Boolean policy ($k=6$). Values visualize the measured dominance of CP-ABE operations in the local prototype.}
\label{fig:breakdown}
\end{figure}

\subsection{Experiment 2: Revocation cost under churn}
Experiment~2 studies the cost of revocation and policy evolution under varying churn, measured as revocations per minute. We compare two approaches: (i) naive per-revocation rekeying, where each revocation triggers immediate $CK$ re-encryption for all protected objects; and (ii) epoch-based rekeying, where $CK$ rotation occurs once per epoch, independent of churn within the epoch.

The artifact's Experiment~2 run uses $M=200$ protected assets over a $T=180$ s window. For epoch length $L=60$ s, $E=3$ epochs; for $L=10$ s, $E=18$ epochs. Under naive per-revocation rekeying, $CK$ updates equal $M \cdot R$, where $R$ is the number of revocations in the window.

Shorter epochs reduce revocation delay but increase cost because more epochs occur per unit time. In our run, epoch=10 s costs 775 s of crypto time over the window, while epoch=60 s costs 129 s. At high churn (10 revokes/min, $R \approx 30$ over 180 s), epoch=60 s is approximately $10\times$ cheaper than naive per-revocation rekeying (129 s versus 1292 s).

\begin{figure}[t]
\centering
\includegraphics[width=0.72\linewidth]{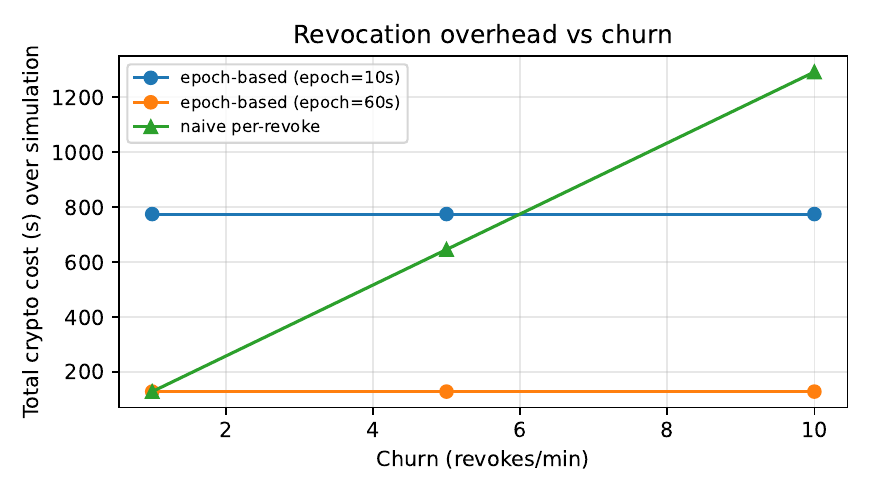}
\caption{Total cryptographic rekeying cost over a 180 s window for naive per-revocation rekeying and epoch-based rekeying with 10 s and 60 s epochs ($M=200$ protected assets).}
\label{fig:revocation}
\end{figure}

\subsection{Experiment 3: Operational scaling with users and assets}
Experiment~3 evaluates how ledger growth and access latency scale with the number of consumers $N$ and the number of protected assets $M$ during a read-heavy workload. The workload generates $M$ objects, appends their metadata records, and then executes read operations from $N$ users under randomly assigned attributes and policies ($k=6$, mixed Boolean).

The artifact run uses $N \in \{50,200\}$ and $M \in \{50,200\}$. Each user performs five reads, so total reads equal $5N$. Because the ledger records per-asset metadata, growth scales with $M$ and is largely independent of $N$. In our runs, increasing $M$ from 50 to 200 increases ledger growth from approximately 309 KB to approximately 1.24 MB, while changing $N$ from 50 to 200 has negligible impact. Median latency increases with $M$ due to a larger working set in the CAS/ledger path, while tail latency remains influenced by CP-ABE decryption attempts.

\begin{figure}[t]
\centering
\includegraphics[width=0.72\linewidth]{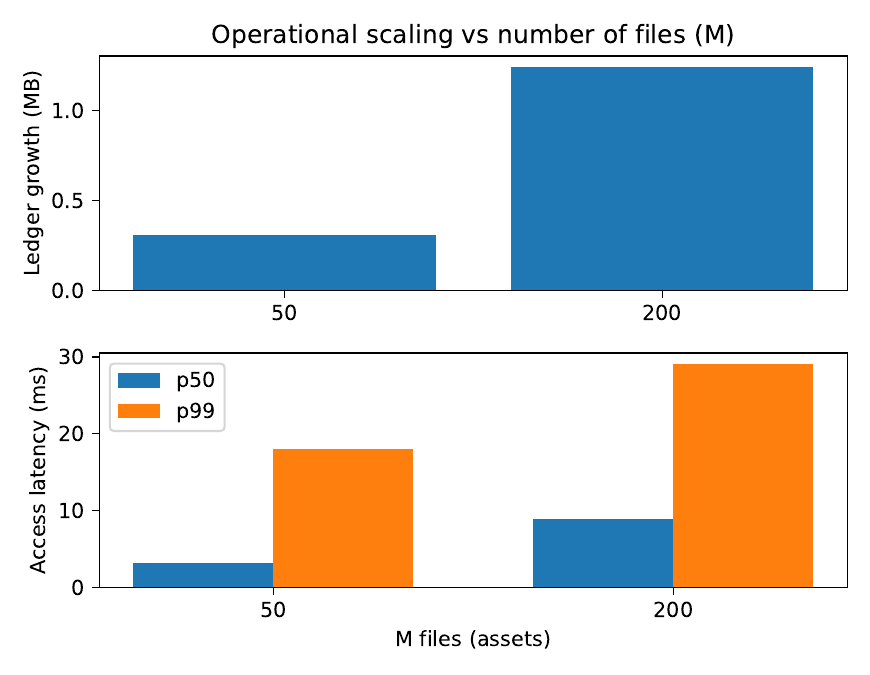}
\caption{Experiment 3 operational scaling: ledger growth and access latency (p50/p99) as the number of protected files $M$ increases, aggregated over user counts $N$.}
\label{fig:scaling}
\end{figure}

\subsection{Experiment 4: Gateway offload trade-off}
Experiment~4 quantifies the benefit of gateway-assisted decryption for resource-bounded clients. We simulate client slow-down factors ($1\times$, $2\times$, $4\times$) and compare the median (p50) end-to-end time when the client performs full decryption versus when decryption is offloaded to a gateway process.

\begin{figure}[t]
\centering
\includegraphics[width=0.72\linewidth]{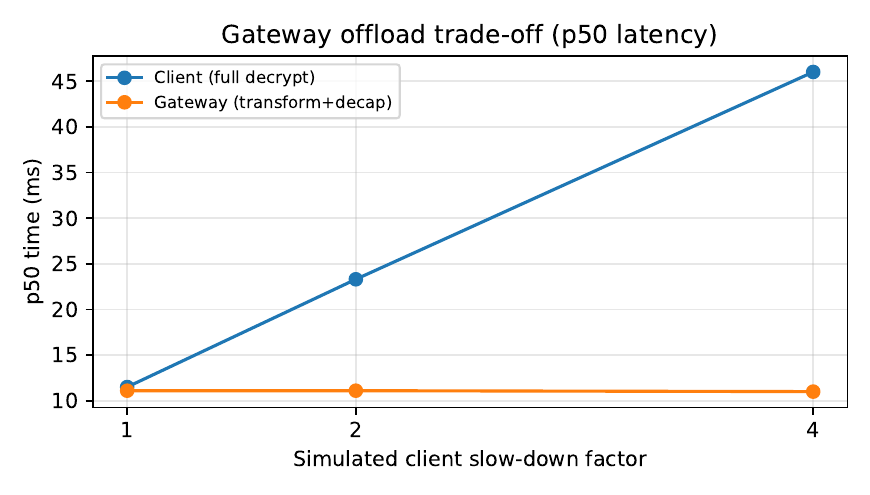}
\caption{Median (p50) time for full client decryption versus gateway-assisted decryption under simulated client slow-down factors. The $4\times$ point is a simulation of resource-constrained client behavior, not a direct measurement on an IoT device.}
\label{fig:gateway}
\end{figure}

As shown in Figure~\ref{fig:gateway}, gateway decryption time remains near-constant at approximately 11 ms across client slow-down factors because the heavy CP-ABE work is performed on the gateway. Client-side p50 grows from 11.5 ms at $1\times$ to 46.0 ms at $4\times$. At $4\times$ slow-down, the gateway mode provides a 4.17$\times$ relief ratio.

\subsection{Experiment 5: Baseline comparison}
Experiment~5 compares our ledger-published ciphertext key approach against a simplified online key server baseline. In the baseline, a trusted server enforces the policy and returns the symmetric key to the requester over an authenticated channel. This bypasses CP-ABE at the client but reintroduces an always-online trusted gate.

\begin{table}[t]
\centering
\caption{Experiment 5 baseline comparison: client latency to obtain a usable AES key (lower is better).}
\label{tab:baseline}
\begin{tabular}{llrrrr}
\toprule
Path & Backend & $n$ & Mean (ms) & p50 (ms) & p99 (ms) \\
\midrule
ledger\_ck & gofe\_fame & 50 & 11.84 & 11.77 & 13.84 \\
online\_key\_server & gofe\_fame & 50 & 1.26 & 1.28 & 1.32 \\
\bottomrule
\end{tabular}
\end{table}

\subsection{Experiment 6: Sensitivity to ledger batching}
Experiment~6 evaluates a simple optimization: batching multiple metadata records into a single ledger append. In production ledgers, such as Fabric, batching and block formation can materially affect throughput~\cite{fabric}. Our prototype ledger is a local hash-chained log, so batching primarily reduces per-append overhead in the application layer.

\begin{figure}[t]
\centering
\includegraphics[width=0.72\linewidth]{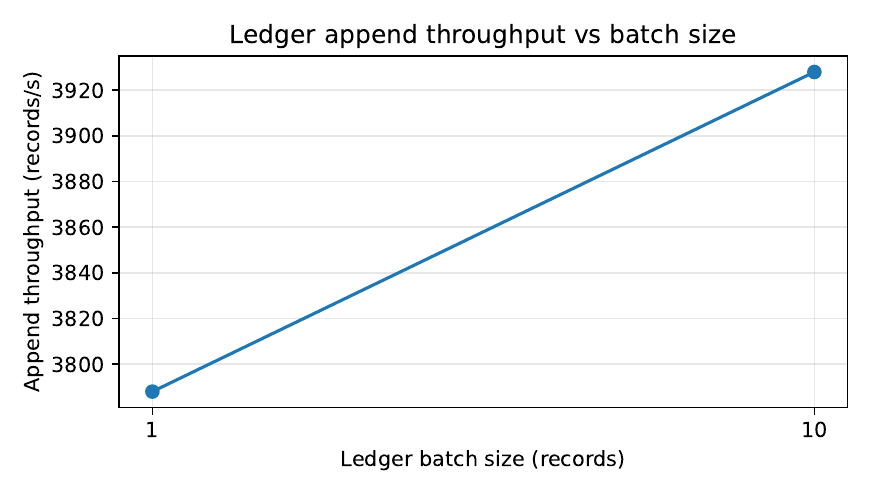}
\caption{Append throughput versus ledger batch size in the local hash-chained log used by the prototype (Experiment 6).}
\label{fig:batching}
\end{figure}

\subsection{Experiment 7: Case study (smart-building maintenance)}
Experiment~7 is a small case study that exercises authorization and revocation semantics in a smart-building maintenance scenario. Three principals, Alice, Bob, and Carl, attempt to decrypt $CK$s protecting maintenance logs. Initially, all three principals hold attributes satisfying the policy and should succeed. Carl is then revoked, an epoch update is performed, and new $CK$s are published.

Before revocation, all principals successfully decrypt 30/30 attempts. After revocation and epoch rollover, Alice and Bob remain authorized (30/30 successes), while Carl's decryption success drops to 0/30, demonstrating effective forward revocation under the epoch mechanism. The case study performs 30 $CK$ updates with a mean $CK$ update time of 58.8 ms. We also include an always-unauthorized outsider, Dana, whose attributes never satisfy the policy; Dana fails both before and after revocation.

\begin{table}[t]
\centering
\caption{Experiment 7 case study: decryption success counts before and after revocation with epoch rollover (30 protected objects).}
\label{tab:case}
\begin{tabular}{lrr}
\toprule
Principal & Before (success/30) & After (success/30) \\
\midrule
alice\_admin & 30/30 & 30/30 \\
bob\_maint & 30/30 & 30/30 \\
carl\_r\_contract & 30/30 & 0/30 \\
dana\_other & 0/30 & 0/30 \\
\bottomrule
\end{tabular}
\end{table}

\section{Discussion}
\paragraph{Security analysis (informal).} Confidentiality of payloads reduces to the confidentiality of AES-GCM~\cite{aesgcm} for the payload and the security of CP-ABE for key encapsulation~\cite{bethencourt,waters,fame}. The ledger and CAS may reveal access patterns and metadata, but they do not reveal $K$ or plaintext without the appropriate attribute key. Auditability is provided by the append-only ledger semantics: the published $CK$s and their associated CIDs form a verifiable history of key releases.

\paragraph{Revocation semantics.} Epoch-based revocation guarantees that a revoked principal loses access after the epoch rolls over and $CK$s are rotated. This provides a tunable bound on exposure, controlled by epoch length, and shifts work from per-revocation events to periodic maintenance. Our evaluation shows that this amortization can substantially reduce total crypto cost under high churn (Experiment~2), but it is not a drop-in replacement for immediate revocation in all settings.

\paragraph{Authority trust and key escrow.} Like most CP-ABE systems, a central authority that issues attribute keys can decrypt any ciphertext, creating key escrow. Deployments can mitigate this via distributed authorities, threshold ABE, or by separating policy attributes among multiple issuers. These extensions are orthogonal to our ledger publication workflow but are not implemented in our prototype.

\paragraph{Policy privacy and metadata leakage.} Publishing $CK$ and policy metadata on a ledger may leak information about policies and user attributes. Techniques such as hidden-policy ABE or storing only policy identifiers/hashes could reduce leakage at additional cost. We treat policy privacy as out of scope.

\paragraph{Prototype limitations.} Our ledger and CAS are local emulations intended to isolate cryptographic and application-layer costs. Absolute throughput will differ under a real Fabric/IPFS deployment due to network, consensus, and block formation effects. Nevertheless, the dominant trends observed in Experiments~1--6 -- ABE dominating cryptographic cost, $M$-driven ledger growth, and gateway relief for resource-bounded clients -- are expected to carry over because they stem from cryptographic complexity and data-model structure rather than implementation details. Real Fabric deployments also add ordering/endorsement and block formation parameters; our local batching sensitivity indicates only modest gains, but the same knob can matter once network and consensus delays dominate.

\paragraph{Owner availability for rekeying.} Epoch rollover by itself is not sufficient: $CK$s must be re-encrypted under the new epoch. If the data owner is offline, a delegated rekey service or gateway with the appropriate capability can perform $CK$ rotation without touching large payloads. This is operationally similar to key rotation services in cloud KMS systems.

\paragraph{Ledger storage overhead.} Each $CK$ is an ABE ciphertext whose size grows with $|P|$. Although our evaluation focuses on latency, the cost model implies that ledger storage grows with $M$ and the number of $CK$ versions, rather than with $N$. In deployments where ledger storage is constrained, policies can be stored by reference ($PolicyID$), and $CK$ size can be reduced by choosing efficient ABE constructions~\cite{fame}.

\section{Conclusion}
We presented a revocation-ready key management layer for blockchain-based IoT data sharing that replaces online key release with ciphertext key publication on a ledger. By combining AES-GCM for bulk data with CP-ABE for key encapsulation and introducing an epoch-based $CK$ rotation mechanism, the design supports expressive ABAC policies and practical forward revocation without re-encrypting large payloads.

Our prototype evaluation demonstrates that CP-ABE dominates store and, to a lesser extent, retrieve latency; that epoch-based revocation amortizes update costs under churn; that gateway-assisted decryption provides more than $4\times$ relief under a simulated $4\times$ client slow-down; and that ledger growth scales with assets rather than readers. These results clarify the operational trade-offs of deploying ABE-based access control in blockchain-IoT systems and provide a reproducible baseline for future work.

\end{document}